\documentclass{style_arxiv}
\usepackage{subfig}

\title{Search for eV Sterile Neutrinos -- The Stereo Experiment}

\ShortTitle{The Stereo Experiment}

\author{\speaker{Julia Haser}\thanks{On behalf of the Stereo collaboration}\\
        \textit{Max-Planck-Institut f\"{u}r Kernphysik, Heidelberg}\\
        \textit{E-mail:} \email{julia.haser@mpi-hd.mpg.de}}

\abstract{In the recent years, major milestones in neutrino physics were accomplished at nuclear reactors: the smallest neutrino mixing angle $\theta_{13}$ was determined with high precision and the emitted antineutrino spectrum was measured at unprecedented resolution. However, two anomalies, the first one related to the absolute flux and the second one to the spectral shape, have yet to be solved. The flux anomaly is known as the Reactor Antineutrino Anomaly and could be caused by the existence of a light sterile neutrino participating in the neutrino oscillation phenomenon. Introducing a sterile state implies the presence of a fourth mass eigenstate, global fits favour oscillation parameters around $\sin^2({2\theta}) \approx 0.09$ and $\Delta m^2 \approx 1\,\mathrm{eV}^2$.\\
The Stereo experiment was built to finally solve this puzzle. It is one of the first running experiments built to search for eV sterile neutrinos and takes data since end of 2016 at ILL Grenoble (France). At a short baseline of 10 metres, it measures the antineutrino flux and spectrum emitted by a compact research reactor. The segmentation of the detector in six target cells allows for measurements of the neutrino spectrum at multiple baselines. An active-sterile flavour oscillation could be unambiguously detected, as it distorts the spectral shape of each cell's measurement differently.\\
This contribution gives an overview on the Stereo experiment, along with details on the detector design, detection principle and the current status of data analysis.}

\FullConference{presented at\\
        EPS-HEP 2017, European Physical Society Conference on High Energy Physics\\
		5-12 July 2017\\
		Venice, Italy}

\begin{document}

\section{Introduction}

Stereo is a reactor antineutrino experiment located at the Institut Laue Langevin (ILL) in Grenoble, France. The experiment addresses the question, whether a light sterile neutrino eigenstate exists, which would correspond to a fourth neutrino mass eigenstate with a mass in the eV range. If so, its participation in the neutrino oscillation phenomenon could explain the missing electron antineutrino flux at short baselines of 100\,m and less, known as the Reactor Antineutrino Anomaly~\cite{Mention:2011rk}. By 2011, global fits of the neutrino oscillation data from experiments all over the world proposed as best fit oscillation parameters values of about $\sin^2({2\theta}) = 0.09$ and $\Delta m^2 = 1.8\,\mathrm{eV}^2$~\cite{Kopp:2013vaa}. Therefore, the Stereo project is searching for sterile neutrinos at short baselines of 9 to 11 metres from a nuclear reactor.
\\
Furthermore, a measurement of the neutrino spectrum produced by ${}^{235}$U is possible at the ILL reactor, giving new input to further understand the spectral shape distortion observed in the neutrino energy spectra measured at commercial reactor cores.

\section{Detector Outline and Sterile Neutrino Search}

\begin{figure}[b]
	\centering
	\includegraphics[width=0.98\textwidth]{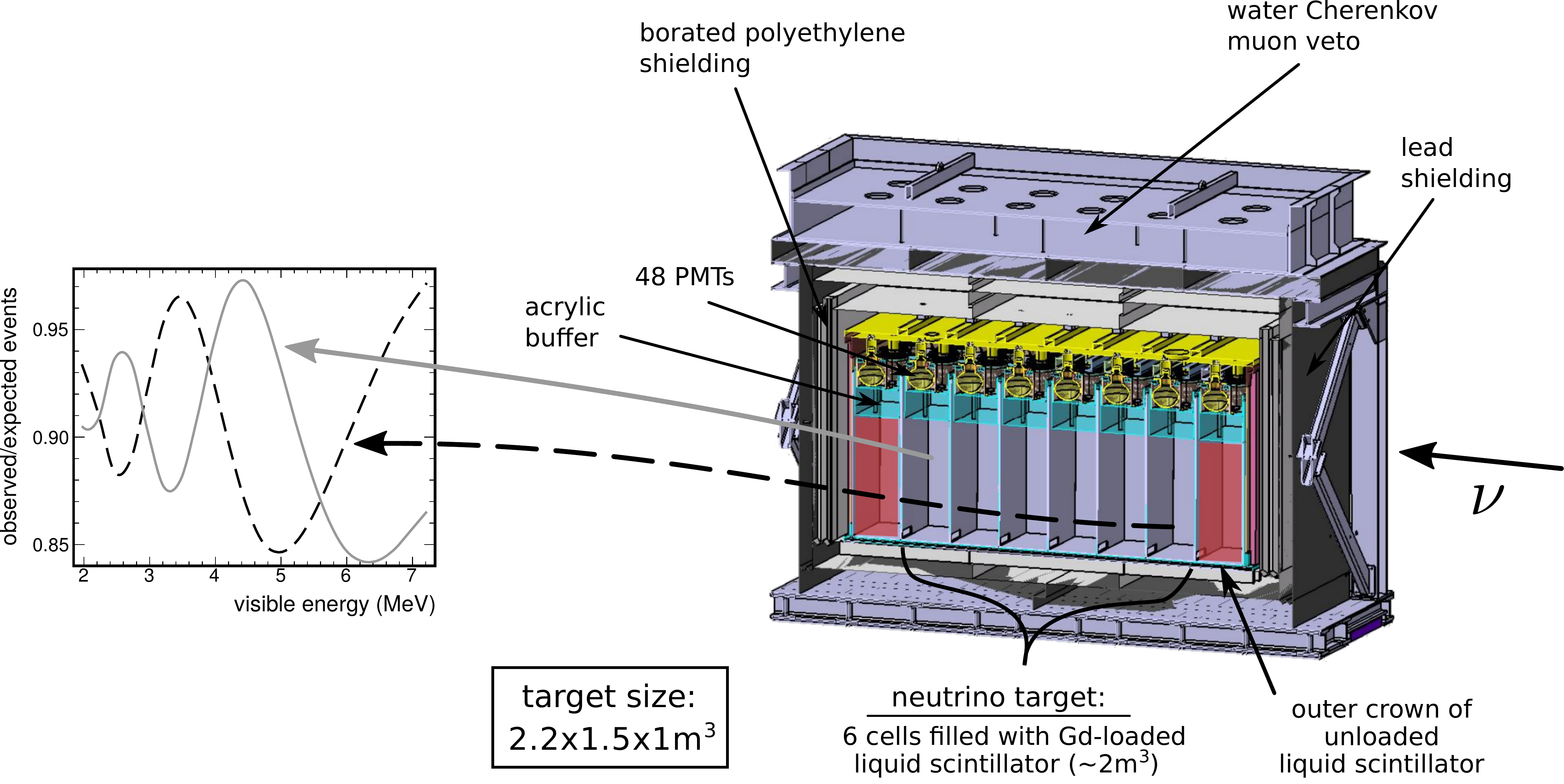}
	\caption{Schematic drawing of the Stereo detector and the principle of sterile neutrino search.}
	\label{fig1}
\end{figure}

\begin{figure}[t]
	\centering
	\includegraphics[width=0.5\textwidth]{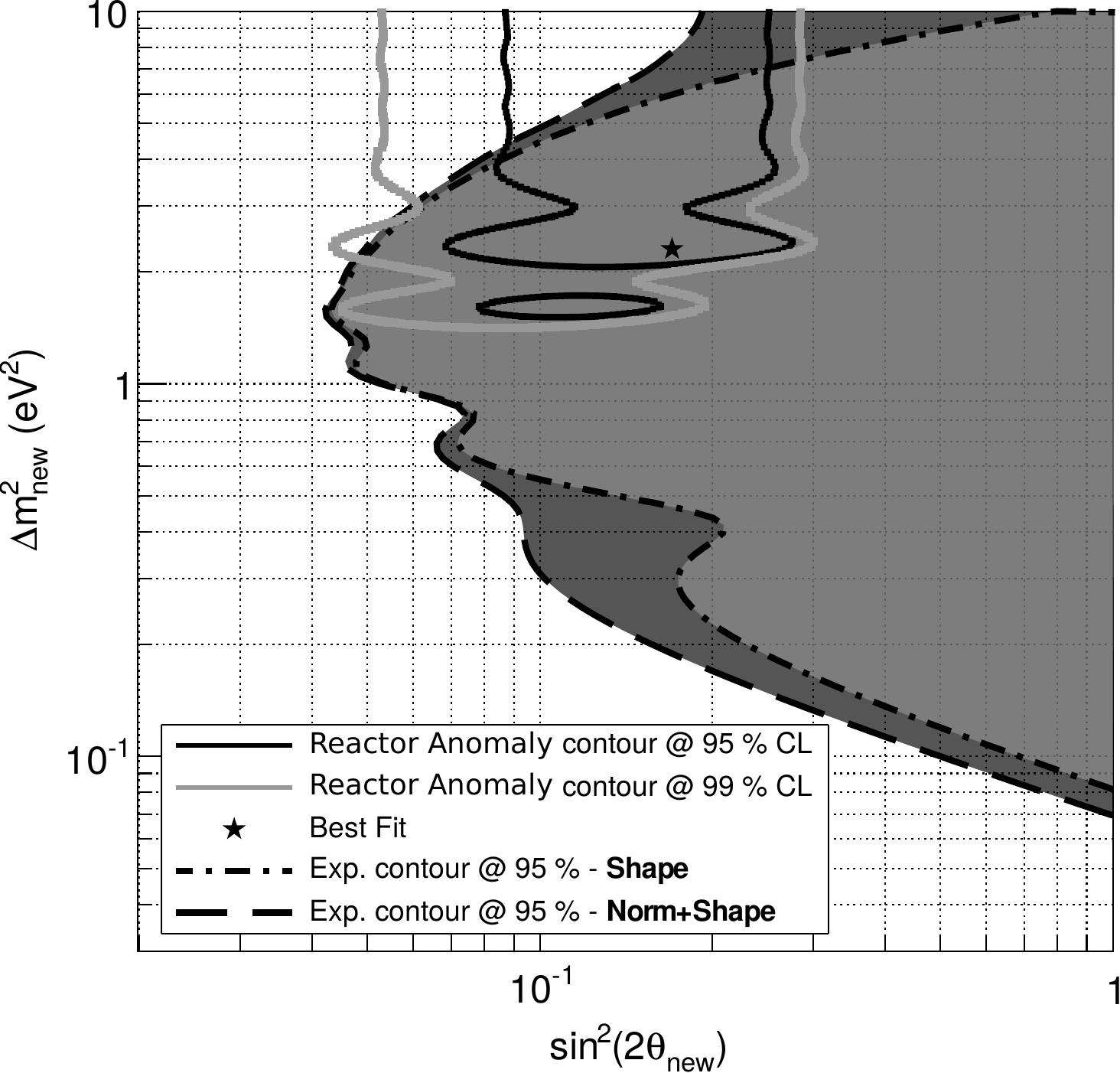}
	\caption{Projected sensitivity of the Stereo experiment for one year reactor-on data.}
	\label{fig2}
\end{figure}

The detector (Fig.\ref{fig1}) measures the neutrino flux and spectrum produced by a compact reactor core of 40\,cm diameter and 80\,cm height. The reactor runs at a thermal power of 58.3\,$\mathrm{MW}_\mathrm{th}$ and with fuel highly enriched in ${}^{235}$U. The electron antineutrinos interact inside the liquid scintillator (LS) of the detector via inverse beta-decay (IBD), which produces a coincidence signal of a positron event followed by a neutron capture event.
The positron kinetic energy is directly linked to the antineutrino energy, its measurement hence allows to reconstruct the incident neutrino spectrum. Secondary particles such as gammas produced by the positron annihilation and the radiative neutron capture create scintillation light in the 2\,$\mathrm{m}^3$ target as well as the scintillator filled outer-crown volume, observed by 48 Hamamatsu PMTs of 8\,inch size.
Both, the LS of target and outer-crown, are admixtures of LAB, PXE and DIN. The latter is added to enhance the pulse-shape discrimination capability for particle identification. Only the target LS is doped with gadolinium in form of a Gd-$\beta$-diketonate complex for efficient neutron detection, confining the fiducial volume to this region. Furthermore, the target is divided along the neutrino path of flight in six equally sized cells of 36\,cm thickness. The cells communicate in terms of liquid exchange, but are optically isolated to the few percent level, allowing for independent measurements of the neutrino spectrum at six different baselines. An oscillation pattern in the relative comparison of these six neutrino spectra would be the proof of active-sterile flavour oscillations with contribution of a $\Delta m^2 \approx 1\,\mathrm{eV}^2$.
\\
After one year of reactor-on data, the Stereo experiment can test a large fraction of the parameter space favoured by the Reactor Antineutrino Anomaly, as shown in the projected sensitivity plot of Fig.\ref{fig2}. The sensitivity study includes the systematic uncertainties from the neutrino reference spectra, a signal-to-background ratio of 1.5 and detector response contributions and uncertainties coming from the energy resolution, energy scale uncertainty and detection efficiency.
\\\\
Since the experimental site is at a shallow depth of 15\,m.w.e.~and as the 58.3\,$\mathrm{MW}_\mathrm{th}$ reactor is mainly used as neutron source by various experiments, heavy shieldings of about 90 tons were installed to keep the backgrounds under control. Shielding materials consist of lead, $\mathrm{B}_4$C, borated polyethylene, soft iron and a mu-metal layer to shield the detector from external magnetic fields.
\\
Cosmic radiation, neutrons and gammas present at the experimental site can mimic the neutrino signature in form of accidental or correlated signal coincidences. The background studies benefit from reactor-off phases. During these periods, a measurement of false neutrino candidates originating only from cosmic rays and remnant internal radioactivity is possible. Further countermeasures against background are the use of a water Cherenkov muon veto above the neutrino detector, a specific set of cuts on the PMT hit pattern of each event, and selections on the topology of the coincidence signal or the pulse-shape discrimination (PSD) parameter.

\section{Calibration and Detector Performance}

\begin{figure}
	\centering
	\captionsetup{width=0.47\textwidth}
	\subfloat[Time stability of the hydrogen capture peak position.]
	{\label{fig3a}\includegraphics[width=0.5\textwidth]{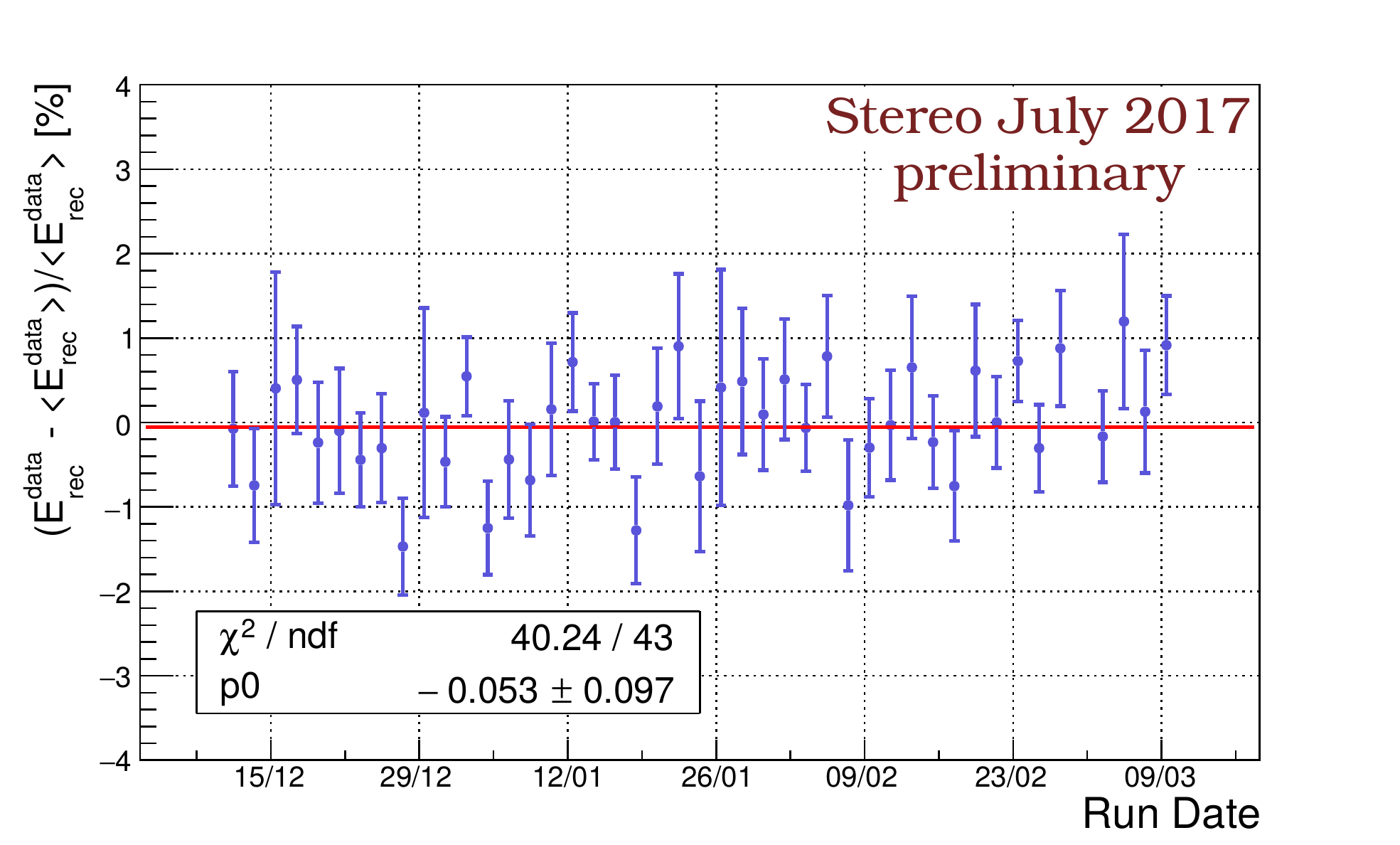}}
	~
	\subfloat[Time stability of the hydrogen capture peak width.]
	{\label{fig3b}\includegraphics[width=0.5\textwidth]{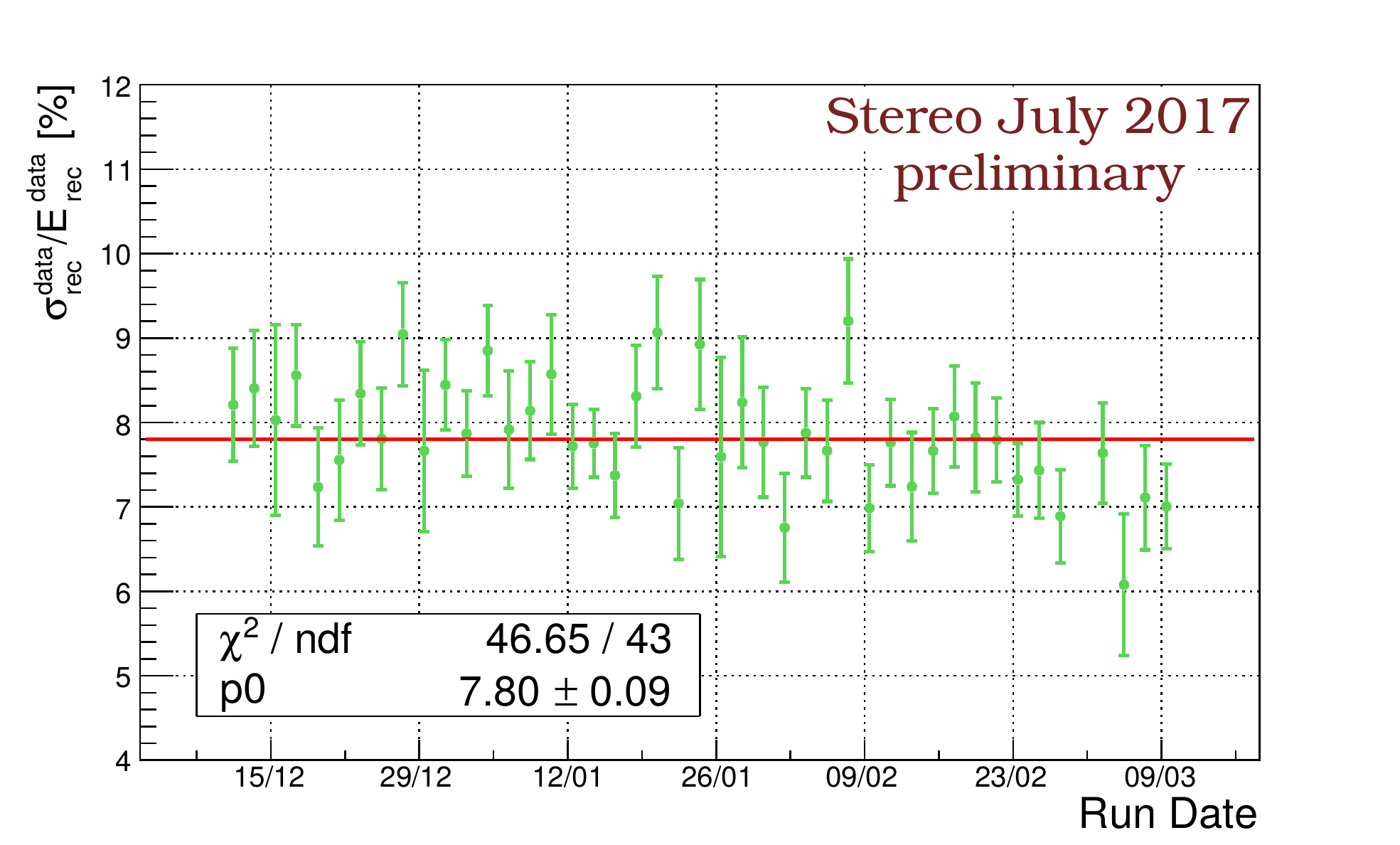}}
	\captionsetup{width=1\textwidth}
	\captionsetup{width=0.9\textwidth}
	\caption{Time stability of the detector response for hydrogen captures of spallation neutrons in the full detector target. The plots show the result of a fit $f(x) = p_0$.}
	\label{fig3}
\end{figure}

\begin{figure}[b]
	\centering
	\includegraphics[width=0.8\textwidth]{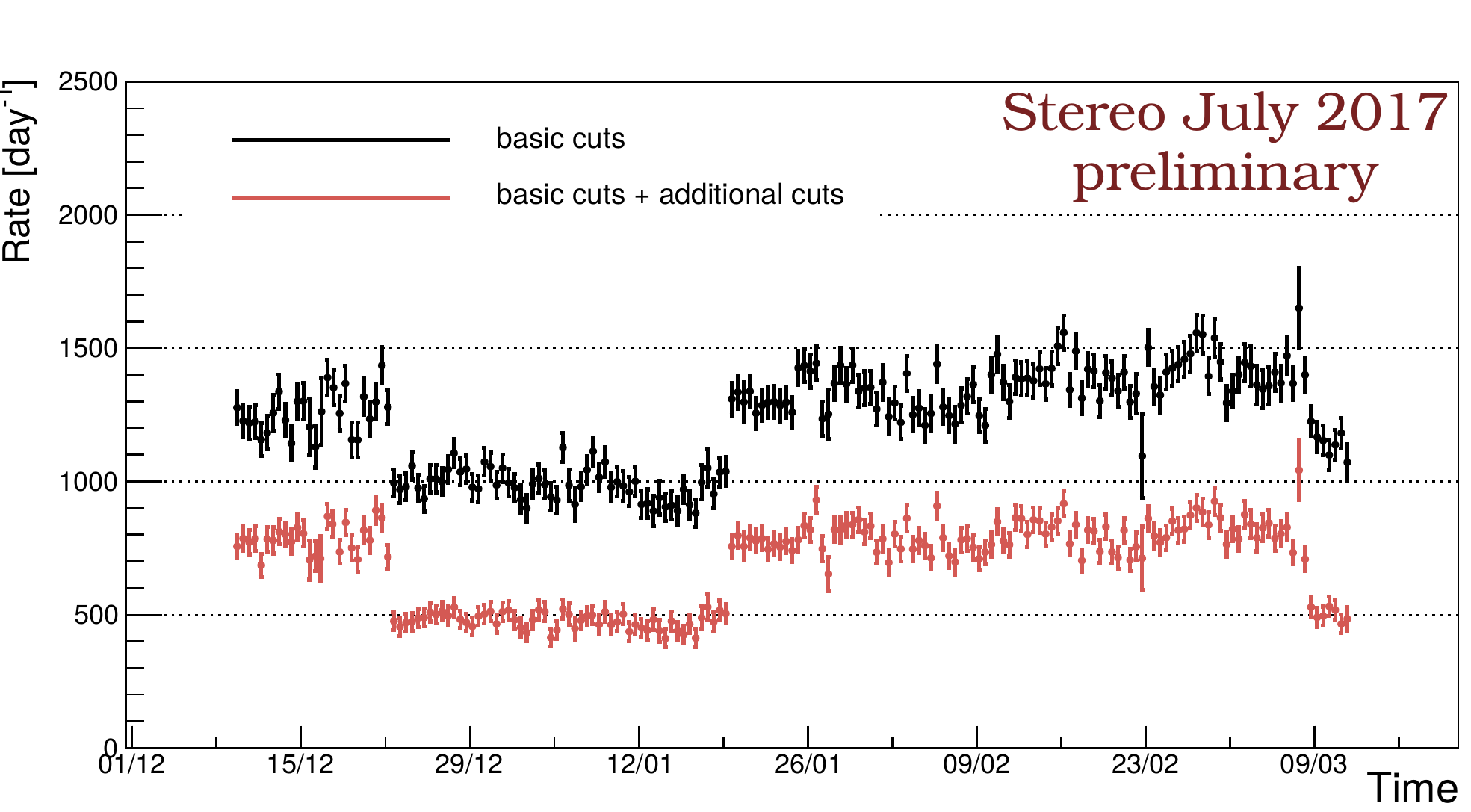}
	\caption{IBD candidate event rate evolution for different selection cuts (see text).}
	\label{fig4}
\end{figure}

The electronics response and the PMT gain are calibrated using an LED light injection system illuminating all target cells and the outer-crown volume. Furthermore, a large variety of radioactive sources can be deployed at various positions inside and outside the detector for calibration purposes. In total, three different calibration systems are available: an internal tube system, an external system able to position sources around the detector and a tube below the detector for calibration from underneath. The internal calibration tubes are used to insert sources into the target cells 1, 4 and 6 at different heights. The external calibration system is installed at the four outer walls of the detector, but inside the shielding, and the source can be moved on rails around the detector at defined horizontal positions.
\\\\
Gamma-neutron sources like AmBe allow for a determination of the detection efficiency of coincidence signals, since neutrons are released simultaneously with a gamma. The latter can be used as starting signal to tag the time of neutron emission. Hence, the capture time of neutrons can be estimated, and is found with $(16.2 \pm 0.2)\,\mu\mathrm{s}$ to be in agreement with the capture time constant of the neutrino candidate events of $(16.5 \pm 0.6)\,\mu\mathrm{s}$. Moreover, the capture time constant of the AmBe neutrons is found to be stable for repeated AmBe source deployments performed over the data taking period. The same calibration data is used to prove an excellent energy containment, as the energy spectrum of the neutron capture events clearly exhibits with good resolution both peaks from radiative neutron captures on hydrogen and gadolinium at 2.2 and 8\,MeV, respectively. Using the delayed spectrum, it is possible to estimate a capture fraction of neutrons on Gd of about 86\,\% at the centre of the target.
\\\\
The set of gamma sources spans energies from 0.5 to 4.4\,MeV and is used for energy scale calibration purposes and to study non-linearities of the response as a function of the deposited energy. The reconstruction algorithm takes into account cell-to-cell variations in acceptance parameters as well as light crosstalk between different cells. Hence, the charge $Q_i$ measured in cell $i$ is understood as a sum of several contributions:
\begin{equation}
Q_i = \alpha_i \displaystyle\sum_{j=\mathrm{cells}} E_j^\mathrm{dep} \times f_j \times L_{j\rightarrow i}\,.
\end{equation}
Here, $E_j^\mathrm{dep}$ is the energy deposited in cell $j$, $f$ the light yield per energy, $L_{j\rightarrow i}$ light crosstalk coefficients describing the leakage of light from cell $j$ to cell $i$ and $\alpha_i$ are cell-dependent light acceptances. All calibration coefficients are obtained from measured data.
\\\\
The detector response was determined to be around 270 PE per MeV. Studies of ${}^{54}$Mn source data show a good homogeneity of the energy response throughout a target cell (3\,\% between the borders and the centre) and agrees with simulation at the percent level. Moreover, the response is stable in time and has a good resolution as seen from the analysis of the hydrogen capture peak position (Fig.\ref{fig3a}) and width (Fig.\ref{fig3b}) from spallation neutron data, which produces events spread throughout the target.
\clearpage
\noindent
The IBD candidate event rate as a function of time is plotted in Fig.\ref{fig4} and clearly shows the stepwise change in the event rate due to reactor-on and -off. The daily rates are corrected for dead time and the dependence on the atmospheric pressure. The black data points show the rate obtained with a basic set of cuts requiring for the total energy of the prompt event $1.5\,\mathrm{MeV} < E < 8\,\mathrm{MeV}$ and less than 1.1\,MeV deposited in the outer-crown. The delayed event is required to have an energy between 5 and 10\,MeV, while more than 1\,MeV has to be detected in the target. Moreover, the prompt and delayed events have to take place in a time period of less than $70\,\mu\mathrm{s}$. For the prompt event signal the PSD parameter is limited to deviate by less than 2.5$\,\sigma$ from the central PSD value found with gamma calibration data or to be lower than that.
\\
Additional cuts exploiting the event topology can be applied on the PMT hit pattern to reduce the amount of background coincidences from stopping muons. The distance between the prompt and the delayed events can be restrained to be less than 0.4\,m along the axis of target segmentation, suppressing a large fraction of accidental coincidences. A further reduction of non-IBD events can be achieved by requiring a valid prompt event to have less than 0.7\,MeV of energy in the cells next to the main cell of the interaction.
Application of these additional cuts on top of the basic set of cuts improves the signal-to-background ratio by a factor of 2, as seen from the red data points in Fig.\ref{fig4}. At the same time, the signal inefficiency due to these additional cuts is on the scale of a few percent.
\\
Furthermore, the heavy shielding is found to be highly efficient: the accidental background was measured to be below the design goal and no major contribution of correlated events by neutrons from the reactor is observed in studies comparing the reactor-on to the reactor-off data.
The signal rate can be determined to be around 300 neutrinos per day. This number meets well the expected neutrino rate.

\section{Status and Outlook}

In a first data taking phase, which started in November 2016, the Stereo experiment collected 70 days of reactor-on and 25 days of reactor-off data. The analysis of this data set is ongoing, including further cut optimisation and more detailed cosmic background studies. After the energy scale reconstruction is finalised, a sterile parameter analysis including the search for spectral shape distortions is foreseen.
\\
With the upcoming data taking period, another 150 days of neutrino data could be recorded by the experiment in 2018.

\end{document}